# Common feature of concave growth pattern of oscillations in terms of speed, acceleration, fuel consumption and emission in car following: experiment and modeling

Junfang Tian

Institute of Systems Engineering, College of Management and Economics, Tianjin University, No. 92 Weijin Road, Nankai District, Tianjin 300072, China, jftian@tju.edu.cn

Rui Jiang

MOE Key Laboratory for Urban Transportation Complex Systems Theory and Technology, Beijing Jiaotong University, Beijing 100044, China, jiangrui@bjtu.edu.cn

Martin Treiber

Technische Universität Dresden, Institute for Transport & Economics, Würzburger Str. 35, D-01062 Dresden, Germany, treiber@vwi.tu-dresden.de

Shoufeng Ma

Institute of Systems Engineering, College of Management and Economics, Tianjin University, No. 92 Weijin Road, Nankai District, Tianjin 300072, China, sfma@tju.edu.cn

Bin Jia, Wenyi Zhang

MOE Key Laboratory for Urban Transportation Complex Systems Theory and Technology, Beijing Jiaotong University, Beijing 100044, China, {bjia@bjtu.edu.cn, wyzhang@bjtu.edu.cn}

This paper has investigated the growth pattern of traffic oscillations by using vehicle trajectory data in a car following experiment. We measured the standard deviation of acceleration, emission and fuel consumption of each vehicle in the car-following platoon. We found that: (1) Similar to the standard deviation of speed, these indices exhibit a common feature of concave growth pattern along vehicles in the platoon; (2) The emission and fuel consumption of each vehicle decrease remarkably when the average speed of the platoon $\bar{v}$ increases from low value; However, when $\bar{v}$ reaches 30*km/h*, the change of emission and fuel consumption with $\bar{v}$ is not so significant; (3), the correlations of emission and fuel consumption with both the standard deviation of acceleration and the speed oscillation are strong. Simulations show that with the memory effect of drivers taken into account, the improved two-dimensional intelligent driver model is able to reproduce the common feature of traffic oscillation evolution quite well.

# 1. Introduction

Traffic oscillations almost happen on highways every day (Treiterer and Myers, 1974; Kühne, 1987; Kerner, 2004; Schönhof and Helbing, 2007; Treiber and Kesting, 2013). Comparing with homogeneous traffic flow, traffic oscillations are undesirable since they cause more fuel consumption, environment pollution, and likely more accidents. In particular, if traffic oscillation amplitude grows large enough, traffic jams will be induced. Some observed features of traffic oscillations are reported. For example: (i) the amplitude of oscillation may grow to some extent when it propagate upstream and then become stable or start to decay (Schönhof and Helbing, 2007; Li and Ouyang, 2011; Zheng et al. 2011); (ii) oscillation exhibits regular periods in the propagation process varying from 2-15*min* (Kühne, 1987; Mauch and Cassidy, 2004; Laval and Leclercq, 2010). The formation of traffic oscillation are attributed to highway bottlenecks, such as highway lane drops (Bertini and Leal, 2005), lane changes near merges and diverges (Laval, 2006; Laval and Daganzo, 2006; Zheng et al. 2011) and roadway geometries (Jin and Zhang, 2005).

To explain the formation and propagation of traffic oscillations and other traffic phenomena, many traffic flow models have been proposed, such as the General-Motors family of car-following models (Chandler et al., 1958; Gazis et al., 1959, 1961), the Lighthill-Whitham-Richards model (Lighthill and Whitham, 1955; Richards, 1956), the Newell model (Newell, 1961), the Payne model (Payne, 1971), the Gipps model (Gipps, 1981), and so on. Traffic engineers usually perform parameter calibration of these models and then compare the simulation results with the empirical data.

A significant change of traffic flow studies occurs in the 1990', represented by the papers of Kerner and Konhäuser (1993, 1994), Bando et al. (1995), Lee et al. (1998, 1999), Helbing et al. (1999), Treiber et al. (2000), and so on. In these models, it is assumed either explicitly or implicitly that there is a unique relationship between speed and spacing in the steady state. Different from previous models, the relationship contains a turning point. As a result, traffic flow might be stable, metastable, or unstable. Disturbances in the metastable and unstable traffic flows could grow and develop into jams via a subcritical Hopf bifurcation (Lee et al., 1998, 1999; Helbing et al., 1999). In the related studies, researchers pay much attention to the spatiotemporal patterns induced by bottlenecks. Several typical patterns have been reported to occur at different bottleneck strength, such as the homogeneous congested traffic (HCT), the oscillating congested traffic (OCT), the triggered stop-and-go traffic (TSG), and so on.

Since traffic flow is classified into free flow and congested flow states in these models, they have been called as two-phase models by Kerner (2013). Although Kerner is one of the pioneers to initial the traffic flow study change in the 1990', he later believes that the theory is not able to correctly simulate real traffic flow (Kerner, 2004; 2013). Kerner proposed the three-phase traffic theory which claims that congested traffic flow should be further classified into synchronized flow and wide moving jam (Kerner and Rehborn, 1996a, 1996b, 1997; Kerner, 1998, 2004, 2009). The synchronized flow occupies a two-dimensional region in the speed-spacing plane in the steady state. A transition from free flow to jam is usually as follows. The phase transition from free flow to synchronized flow, which is caused by the discontinuous characteristic of the probability of over-acceleration, occurs firstly. The emergence of wide moving jams from synchronized flow occurs later and at different locations. Three-phase traffic theory believes that in general situations, wide moving jams can emerge only in synchronized flow. The direct transition from free flow to wide moving jams happens only if the formation of synchronized flow is strongly hindered due to a non-homogeneity, in particular at a traffic split on a highway (Kerner, 2000). Typical patterns that occur at different bottleneck strength include: general pattern (GP, which can be regarded as synchronized pattern + traffic jams), widening synchronized flow pattern (WSP), moving synchronized flow pattern (MSP). These are also different from the patterns in two-phase models.

The debate between two-phase models and three-phase theory is on-going (Schönhof and Helbing, 2007, 2009; Helbing et al, 2009; Treiber et al., 2010; Kerner, 2013). Traffic researchers realized that this is due to a lack of high-fidelity trajectory data that fully cover the evolution of the congestions. The NGSIM-data (NGSIM, 2006) and the helicopter data (Ossen et al., 2006) are two such efforts. However, unfortunately, both data sets cover only several hundred meters of the highway and contain many confounding factors.

A recent effort is reported in Jiang et al. (2014, 2015), in which an experimental study of car following behaviors in a 25-car-platoon on an open road section has been conducted. They found that the standard deviation of the time series of the speed of each car increases in a concave way along the platoon. They showed that this feature contradicts the simulation results of two-phase models. However, if traffic states are allowed to span a two-dimensional space as supposed in the three-phase traffic theory, the concave growth of speed oscillation can be qualitatively or quantitatively reproduced.

In the previous papers, only the standard deviation of speed has been measured. This paper makes a further analysis of the traffic oscillation feature. We study the standard deviation of acceleration, the fuel consumption and the emission in the traffic oscillations, which are critical to monitor traffic performance and evaluate highway services (Li, et al. 2014). It has been found these three indices exhibit a common feature of concave growth along the platoon as speed oscillations. These findings reveal the connections between car-following behaviors of individual drivers and the growth pattern of oscillation along the platoon, and thus have significant implications to car-following modeling.

The paper is organized as follows. Section 2 firstly briefly reviews the experimental setup and previous experimental results. Then new experimental results are presented. Section 3 performs the simulations of a car following model. Although simulation results of speed oscillation agree with the experimental ones, the quantitative deviation between simulation results of standard deviation of acceleration, emission and fuel consumption and experimental ones is remarkable. Section 4 introduces memory effect of the drivers into the model. Simulation results of the modified model quantitatively agree better with the experimental ones. Section 5 concludes the paper.

**2 Experimental setup and results**

*2.1 Experimental setup and previous results*

Jiang et al. (2014, 2015) have conducted a 25-car-platoon experiment on a 3.2*km* non-signalized stretch in a suburban area in Hefei City, China. High-precision differential GPS devices were installed on all of the cars to record their locations and velocities every 0.1*s*. During the experiment, the driver of the leading car is asked to drive the car at certain pre-determined constant speed. Other drivers in the experiment are required to drive their cars as they normally do, following each other without overtaking. When reaching the end of the road section, the car platoon decelerates, makes U-turn, and stops. When all the cars have stopped, a new run of the experiment begins.

It has been found that:

(i) Even if the preceding car moves with an essentially constant speed, the spacing of the following car still fluctuates significantly; drivers could adopt a smaller spacing traveling at a higher speed than that traveling at a lower speed; the fluctuation in spacing as well as the average spacing between vehicles can be significantly different while their average speed is almost the same; the length of the platoon can differ sizably even if the average speed of the platoon is essentially the same.

(ii) Stripe structure has been observed in the spatiotemporal evolution of traffic flow, which corresponds to the formation and development of oscillations. When traffic flow speed is low, cars in the rear part of the 25-car-platoon will move in a stop-and-go pattern.

(iii) The standard deviation of the time series of the speed of each car increases along the platoon in a concave or linear way, see Fig.1. However, the physical limits of speeds implies that if we had a much longer platoon, the variations of speed of cars in the tail of the platoon would be capped and the line would bend downward, making the overall curve concave shaped.

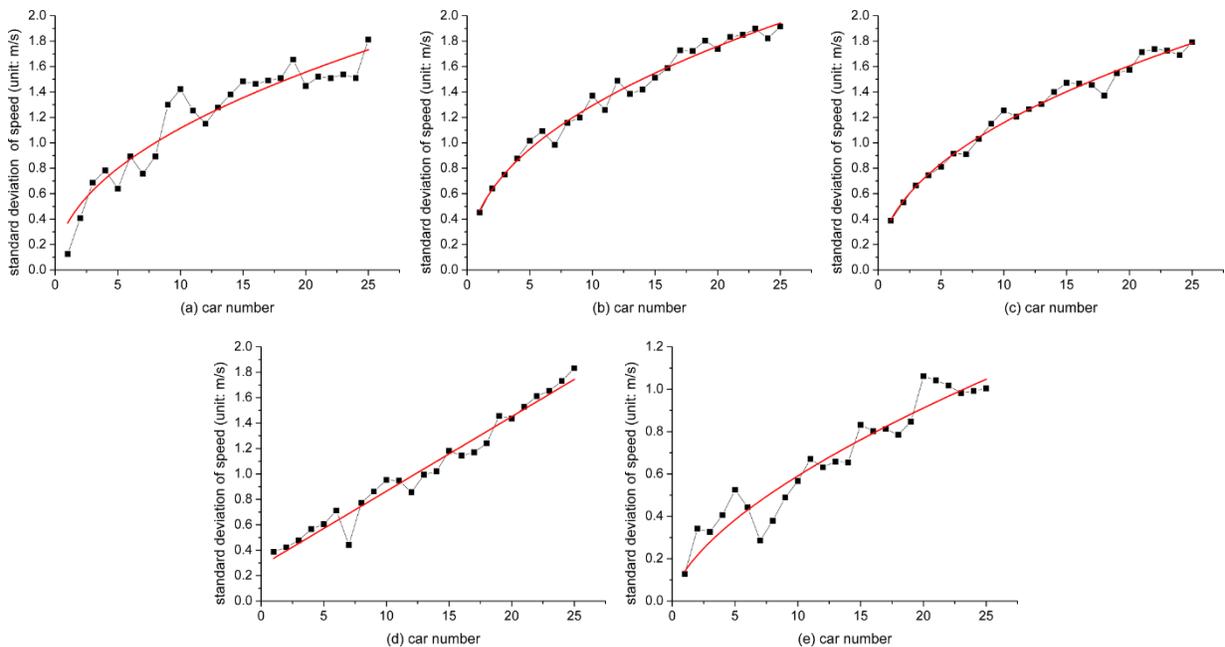

**Fig. 1.** The standard deviation of the time series of the speed of each car in the car following experiments. The symbol solid black lines

are the experiment results and the red lines are the fitted lines. From (a) to (e), the leading car moves with $v_{leading}$ =50, 40, 30, 15, 7$km/h$ respectively. The car number 1 is the leading car.

Jiang et al. (2014, 2015) have shown that the simulation results of traditional car following models, such as the General Motor models (GMs, Chandler et al., 1958; Gazis et al. 1961; Edie, 1961), Gipps' Model (Gipps, 1981), Optimal Velocity Model (OVM, Bando et al., 1995), Full Velocity Difference Model (FVDM, Jiang et al., 2001) and Intelligent Driver Model (IDM, Treiber et al., 2000), run against these experimental findings. In these models, the standard deviation initially increases in a convex way in the unstable density range. Based on these observations, they have proposed two possible mechanisms to produce this feature. (i) At a given speed, drivers do not have a fixed preferred spacing. Instead they change their preferred spacing either intentionally or unintentionally from time to time in the driving process. (ii) In a certain range of spacing, drivers are not so sensitive to the changes in spacing when the speed differences between cars are small. Only when the spacing is large (small) enough, will they accelerate (decelerate) to decrease (increase) the spacing. Models have been proposed based on the two mechanisms, which were shown to reproduce the experimental findings quite well.

*2.2 Calculation of acceleration*

This paper makes a further analysis of the traffic oscillation feature. Apart from the standard deviation of speed, we study the standard deviation of acceleration, the fuel consumption and the emission in the traffic oscillations. To this end, we need to calculate the acceleration of each car, which is needed to calculate not only the acceleration standard deviation but also fuel consumption and emissions. To determine fuel consumption and emissions, we will apply the VT-Micro model, in which speed and acceleration are two input variables.

Since the GPS devices record the speed every $\Delta t = 0.1s$, we calculate the acceleration via

$$a_n(t) = \frac{v_n(t) - v_n(t - \Delta t)}{\Delta t} \qquad (1)$$

where $a_n(t)$ and $v_n(t)$ is the acceleration and speed of vehicle *n* at time *t*. Fig.2(a) shows an example of the acceleration time series calculated via Eq.(1). One can see that the random fluctuations are very strong. To reduce them, we employ the moving average method. Fig.2(b) and (c) show the smoothed time series for different time windows. One can see that, for a time window of 5 data points, i.e., 0.5$s$, the random fluctuations are still observable. When the time window is equal to 1$s$, the random fluctuations are basically eliminated. When the time window reaches 2$s$, the true acceleration peaks have been damped. As a result, we choose a time window of 1$s$ to calculate the acceleration. We have examined the results by changing time window in the range between 0.5$s$ and 2$s$, and found only minor quantitative differences in the results.

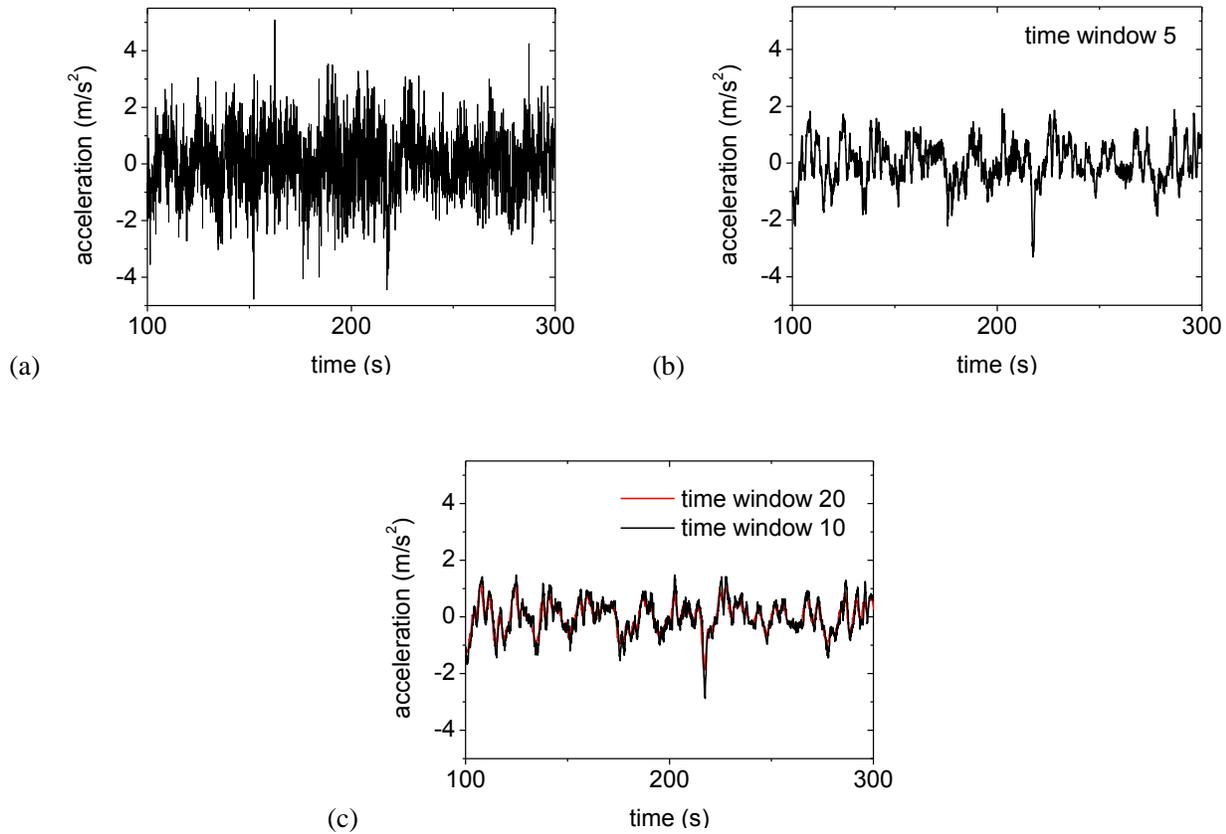

**Fig. 2.** The calculation of acceleration.

## 2.3 Calculation of fuel consumption and emission

To calculate fuel consumption and emission, there are many existing models (such as Ferreira, 1985; Barth et al., 2000; Rakha et al., 2011; Koupal et al., 2002; Hausberger et al., 2003; Wu et al., 2011). We here employ the VT-Micro model (Ahn, 1998; Ahn and Aerde, 2002) which has reasonable estimation accuracy and was validated with field data. Note that this kind of consumption/emission model belongs to the class "modal consumption/emission model" with the subclass "regression-based modal models". There is also the "physics based modal models". For more details, see Chapter 20 of Treiber and Kesting (2013).

The VT-Micro model is given with the exponential function that is called as the measure of effectiveness (*MOE*),

$$MOE\left(a_n(t), v_n(t)\right) = e^{P(a_n(t), v_n(t))} \tag{2}$$

where the exponent *P* is a polynomial function of speed and acceleration,

$$P\left(a_n(t), v_n(t)\right) = \sum_{i=0}^{3} \sum_{j=0}^{3} K_{ij} \left(v_n(t)\right)^i \left(a_n(t)\right)^j \tag{3}$$

The coefficients $K_{ij}$ are regression coefficients from field measurements. Ahn et al. (2002) used their experimental data collected at the Oak Ridge National Laboratory to calibrate the corresponding regression coefficients $K_{ij}$ of the vehicle's fuel consumption, $CO_2$ emission and $NO_x$ emission and obtained the corresponding regression coefficients (see Table 1-3).

**Table 1.** Coefficients for the MOE of fuel consumption (the units of fuel consumption, speed and acceleration are in *liters*/*s*, *km/h*, and *km/h/s*, respectively)

| $K_{ij}$ | $a_n(t)$ is positive | | | | $a_n(t)$ is negative | | | |
|---|---|---|---|---|---|---|---|---|
| | $j = 0$ | $j = 1$ | $j = 2$ | $j = 3$ | $j = 0$ | $j = 1$ | $j = 2$ | $j = 3$ |
| $i = 0$ | -7.735 | 0.2295 | -5.61E-03 | 9.77E-05 | -7.735 | -0.01799 | -4.27E-03 | 1.88E-04 |
| $i = 1$ | 0.02799 | 0.0068 | -7.72E-04 | 8.38E-06 | 0.02804 | 7.72E-03 | 8.38E-04 | 3.39E-05 |
| $i = 2$ | -2.23E-04 | -4.40E-05 | 7.90E-07 | 8.17E-07 | -2.20E-04 | -5.22E-05 | -7.44E-06 | 2.77E-07 |
| $i = 3$ | 1.09E-06 | 4.80E-08 | 3.27E-08 | -7.79E-09 | 1.08E-06 | 2.47E-07 | 4.87E-08 | 3.79E-10 |

**Table 2.** Coefficients for the MOE of $CO_2$ emission (the units of emission, speed and acceleration are in *mg/s*, *km/h*, and *km/h/s*, respectively).

| $K_{ij}$ | $a_n(t)$ is positive | | | | $a_n(t)$ is negative | | | |
|---|---|---|---|---|---|---|---|---|
| | $j = 0$ | $j = 1$ | $j = 2$ | $j = 3$ | $j = 0$ | $j = 1$ | $j = 2$ | $j = 3$ |
| $i = 0$ | 6.916 | 0.217 | 2.35E-04 | -3.64E-04 | 6.915 | -0.032 | -9.17E-03 | -2.89E-04 |
| $i = 1$ | 0.02754 | 9.68E-03 | -1.75E-03 | 8.35E-05 | 0.0284 | 8.53E-03 | 1.15E-03 | -3.06E-06 |
| $i = 2$ | -2.07E-04 | -1.01E-04 | 1.97E-05 | -1.02E-06 | -2.27E-04 | -6.59E-05 | -1.29E-05 | -2.68E-07 |
| $i = 3$ | 9.80E-07 | 3.66E-07 | -1.08E-07 | 8.50E-09 | 1.11E-06 | 3.20E-07 | 7.56E-08 | 2.95E-09 |

**Table 3.** Coefficients for the MOE of $NO_x$ emission (the units of emission, speed and acceleration are in *mg/s*, *km/h*, and *km/h/s*, respectively).

| $K_{ij}$ | $a_n(t)$ is positive | | | | $a_n(t)$ is negative | | | |
|---|---|---|---|---|---|---|---|---|
| | $j = 0$ | $j = 1$ | $j = 2$ | $j = 3$ | $j = 0$ | $j = 1$ | $j = 2$ | $j = 3$ |
| $i = 0$ | -1.08 | 0.2369 | 1.47E-03 | -7.82E-05 | -1.08 | 0.2085 | 2.19E-02 | 8.82E-04 |
| $i = 1$ | 1.79E-02 | 4.05E-02 | -3.75E-03 | 1.05E-04 | 2.11E-02 | 1.07E-02 | 6.55E-03 | 6.27E-04 |
| $i = 2$ | 2.41E-04 | -4.08E-04 | -1.28E-05 | 1.52E-06 | 1.63E-04 | -3.23E-05 | -9.43E-05 | -1.01E-05 |
| $i = 3$ | -1.06E-06 | 9.42E-07 | 1.86E-07 | 4.42E-09 | -5.83E-07 | 1.83E-07 | 4.47E-07 | 4.57E-08 |

*2.4 New experimental results*

Fig.3 and 4 show the standard deviation of acceleration, fuel consumption and emission of each car along the platoon, respectively. It can be seen that: (i) Similar to the speed oscillation, these indices exhibit a common feature of concave growth way along vehicles in the platoon; (ii) As the average speed of the platoon $\bar{v}$ (which equals to $v_{leading}$) decreases, the growth pattern of emission and fuel consumption is more and more close to a linear way; (iii) The emission and fuel consumption of each car decrease remarkably when $\bar{v}$ increases from 7$km/h$ to 15$km/h$, see Fig.5. When $\bar{v}$ further increases from 15$km/h$ to 30$km/h$, fuel consumption and emission of $CO_2$ also decrease remarkably. When $\bar{v}$ continues to increase, the fuel consumption and emission of $CO_2$ only slightly decreases. For emission of $NO_x$, the dependence on $\bar{v}$ is not so regular. Nevertheless, roughly speaking, the change of emission of $NO_x$ is not so significant when $\bar{v}$ increases from 15$km/h$. Finally, Table 4 shows that the correlations of emission and fuel consumption with both the speed oscillation and the standard deviation of acceleration are strong.

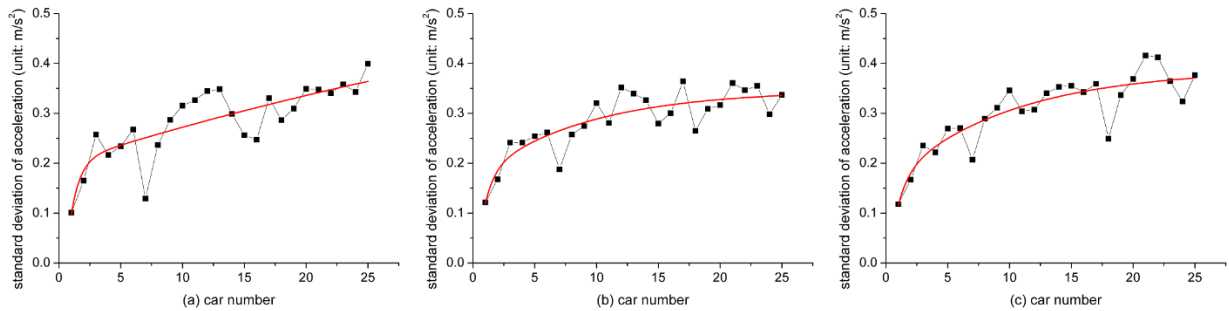

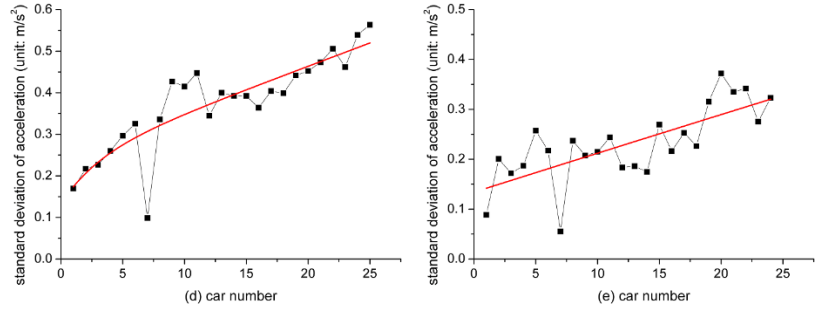

**Fig. 3.** The standard deviation of the time series of the acceleration of each car in the car following experiments. The symbol solid black lines are the experiment results and the red lines are the fitted lines. From (a) to (e), the speed of the leading car moves with $v_{leading}$ =50, 40, 30, 15, 7$km/h$ respectively. The car number 1 is the leading car.

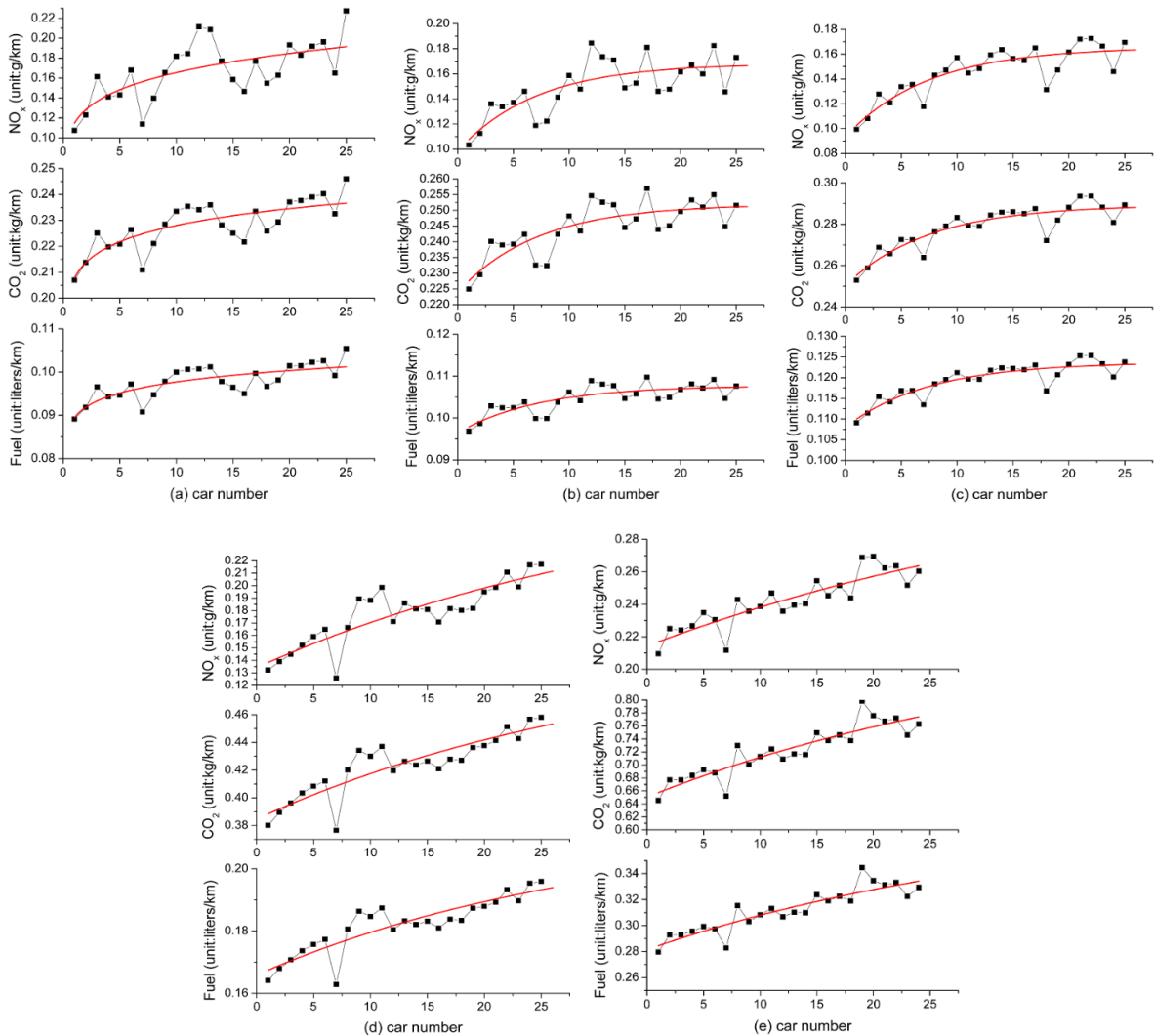

**Fig. 4.** The emission and fuel consumption of each car in the platoon. The symbol solid black lines are the experiment results and the red

lines are the fitted lines. From (a) to (e), the speed of the leading car moves with $v_{\text{leading}}$ =50, 40, 30, 15, 7$km/h$ respectively. The car number 1 is the leading car.

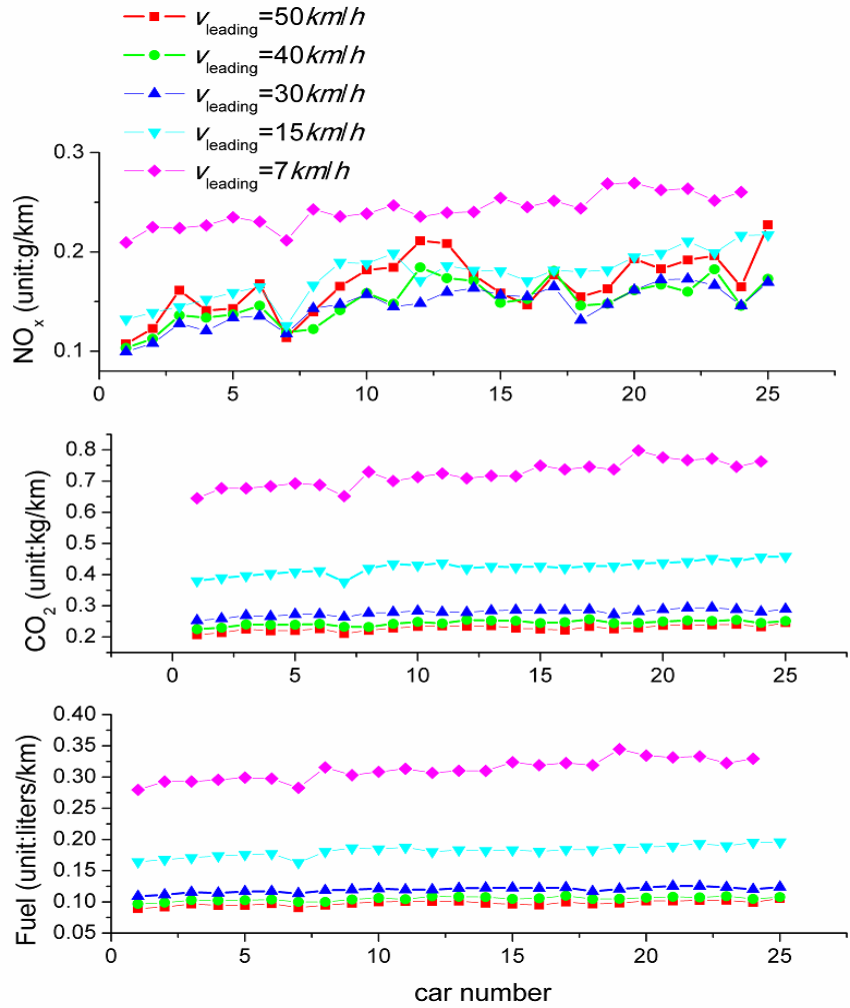

**Fig. 5.** The emission and fuel consumption of each car in the platoon with different $v_{\text{leading}}$.

**Table 4.** Experimental and simulation results of the correlations of emission and fuel consumption with the speed oscillation and the standard deviation of acceleration. $\rho_{v,\text{NOx}}$, $\rho_{v,\text{CO2}}$ and $\rho_{v,\text{Fuel}}$ are the correlations between the speed oscillation and NOx emission, CO$_2$ emission, and fuel consumption, respectively. $\rho_{a,\text{NOx}}$, $\rho_{a,\text{CO2}}$ and $\rho_{a,\text{Fuel}}$ are the correlations between standard deviation of acceleration and NOx emission, CO$_2$ emission, and fuel consumption, respectively.

| $v_{\text{leading}}$ | 50 | | 40 | | 30 | | 15 | | 7 | |
|---|---|---|---|---|---|---|---|---|---|---|
| ($km/h$) | Experi | 2D-IDMM | Experi | 2D-IDMM | Experi | 2D-IDMM | Experi | 2D-IDMM | Experi | 2D-IDMM |
| $\rho_{v,NOx}$ | 0.71 | 0.99 | 0.77 | 0.99 | 0.89 | 1.00 | 0.90 | 0.99 | 0.93 | 1.00 |
| $\rho_{a,NOx}$ | 0.94 | 0.96 | 0.95 | 0.96 | 0.99 | 0.96 | 0.99 | 0.97 | 0.93 | 0.95 |
| $\rho_{v,CO2}$ | 0.81 | 0.99 | 0.82 | 0.99 | 0.92 | 0.97 | 0.92 | 0.98 | 0.92 | 0.99 |
| $\rho_{a,CO2}$ | 0.98 | 0.96 | 0.97 | 0.97 | 1.00 | 0.99 | 0.99 | 0.97 | 0.89 | 0.96 |
| $\rho_{v,Fuel}$ | 0.80 | 0.99 | 0.82 | 0.99 | 0.92 | 0.99 | 0.92 | 0.98 | 0.92 | 0.99 |
| $\rho_{a,Fuel}$ | 0.98 | 0.96 | 0.96 | 0.97 | 1.00 | 0.97 | 0.99 | 0.97 | 0.88 | 0.96 |

## 3 Simulations results of the 2D-IIDM

This section reports simulation results of the improved two-dimensional IDM (2D-IIDM, Tian et al. 2016), which can simulate the synchronized flow and the concave growth pattern of the speed oscillation quite well. The 2D-IIDM is given by,

If $d_{n,\text{desired}}(t) \leq d_n(t)$

$$a_n(t) = a_{\max}\left(1-\left(\frac{v_n(t)}{v_{\max}}\right)^4\right)\left(1-\left(\frac{d_{n,\text{desired}}(t)}{d_n(t)}\right)^2\right)$$

else

$$\begin{cases} \text{If } v_n(t) \leq v_c \\ \quad a_n(t) = a_{\max}\left(1-\left(\frac{d_{n,\text{desired}}(t)}{d_n(t)}\right)^2\right) \\ \text{else} \\ \quad a_n(t) = \min\left(a_{\max}\left(1-\left(\frac{d_{n,\text{desired}}(t)}{d_n(t)}\right)^2\right), -b\right) \end{cases} \quad (4)$$

where $v_c$ is the critical speed, $b$ is the comfortable deceleration and $a_{\max}$ is the maximum acceleration. $d_n(t)$ is the spacing between vehicle $n$ and its preceding vehicle $n+1$, $d_n(t) = x_{n+1}(t) - x_n(t) - L_{\text{veh}}$, $x_n(t)$ is the position of vehicle $n$ and $L_{\text{veh}}$ is the length of the vehicle. $d_{n,\text{desired}}(t)$ is the desired space gap:

$$d_{n,\text{desired}}(t) = \max\left(v_n(t)T(t) - \frac{v_n(t)\Delta v_n(t)}{2\sqrt{a_{\max}b}}, 0\right) + d_0 \quad (5)$$

where and $d_0$ is the jam gap. $T(t)$ is the desired time gap:

$$T(t+\Delta t) = \begin{cases} T_1 + rT_2 & \text{if } r_1 < p_1 \text{ and } v_n(t) \leq v_c, \\ T_3 + rT_4 & \text{if } r_1 < p_2 \text{ and } v_n(t) > v_c, \\ T(t) & \text{otherwise.} \end{cases} \quad (6)$$

where $r$ and $r_1$ are two independent random numbers between 0 and 1. The parameters $T_1$, $T_2$, $T_3$ and $T_4$ indicating the range of the time gap variations give rise to two-dimensional flow-density data in congested states and the two-dimensional region in the flow-density plane is divided into two different sub-regions by the critical speed $v_c$.

In the simulation, the parameters are set as: $v_{max}=30m/s$, $a_{max}=0.8m/s^2$, $b=1.5m/s^2$, $d_0=1.5m$, $v_c=14m/s$, $T_1=0.5s$, $T_2=1.9s$, $T_3=0.9s$, $T_4=1.5s$, $p_1=0.015s^{-1}$, $p_2=0.015s^{-1}$, $\Delta t=0.1s$, and $L_{veh}=5m$. Fig.6 shows that the concave growth of speed oscillation agrees pretty well with the experimental results. Nevertheless, although the concave growth pattern of the standard deviation of acceleration, emission and fuel consumption can be qualitatively simulated, the quantitative deviation between simulation results and experimental ones is remarkable, see Fig.7 and 8.

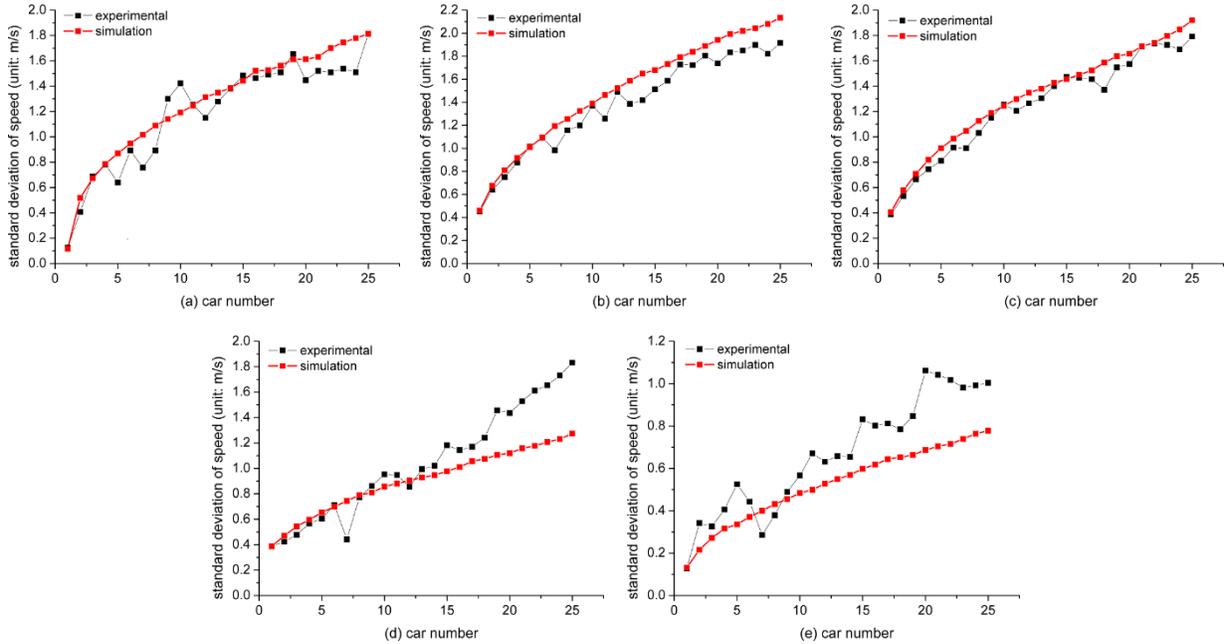

**Fig. 6.** The standard deviation of the time series of the speed of each car. The symbol solid black lines are the experiment results and the symbol solid red lines are the simulation results. From (a) to (e), the speed of the leading car moves with $v_{leading}$=50, 40, 30, 15, 7$km/h$ respectively. The car number 1 is the leading car.

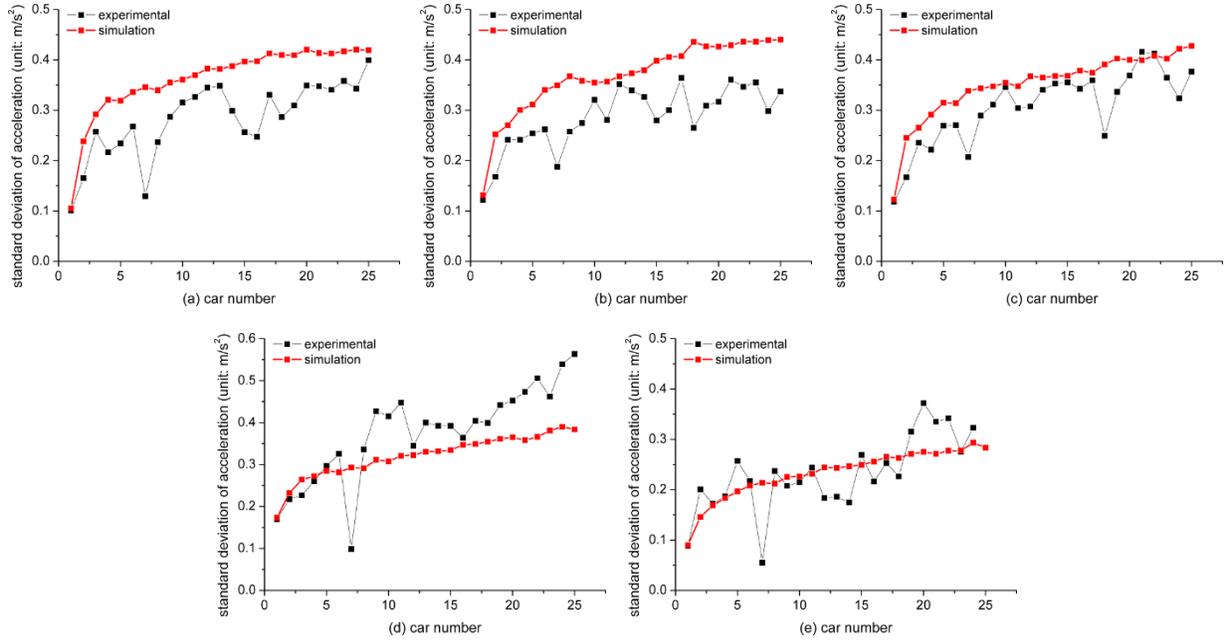

**Fig. 7.** The standard deviation of the time series of the acceleration of each car in the car following experiments. The symbol solid black lines are the experiment results and the red lines are the fitted lines. From (a) to (e), the speed of the leading car moves with $v_{leading}$ =50, 40, 30, 15, 7 $km/h$ respectively. The car number 1 is the leading car.

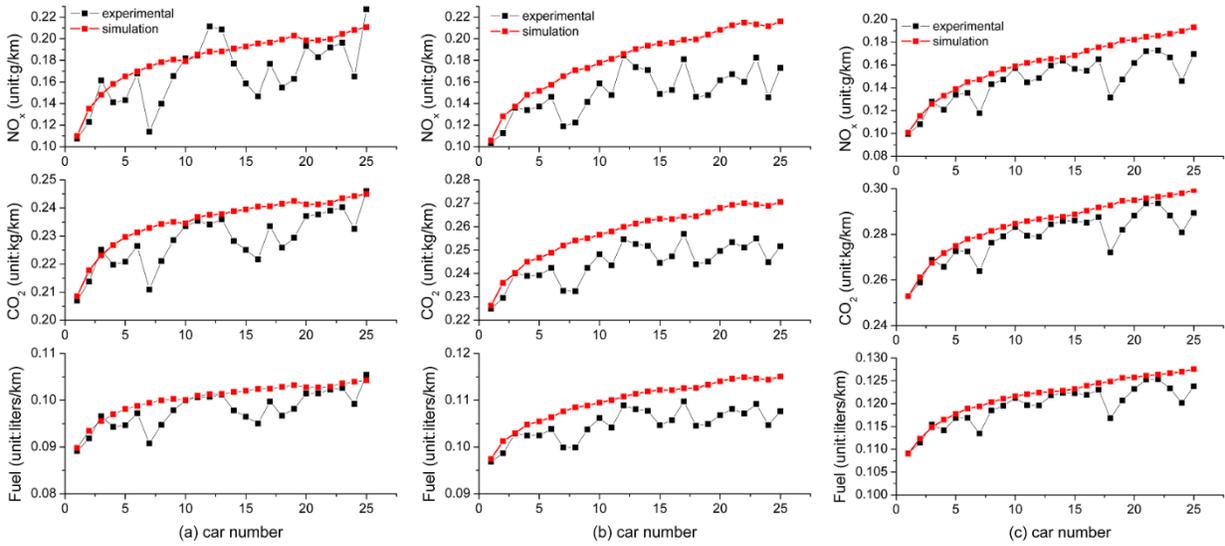

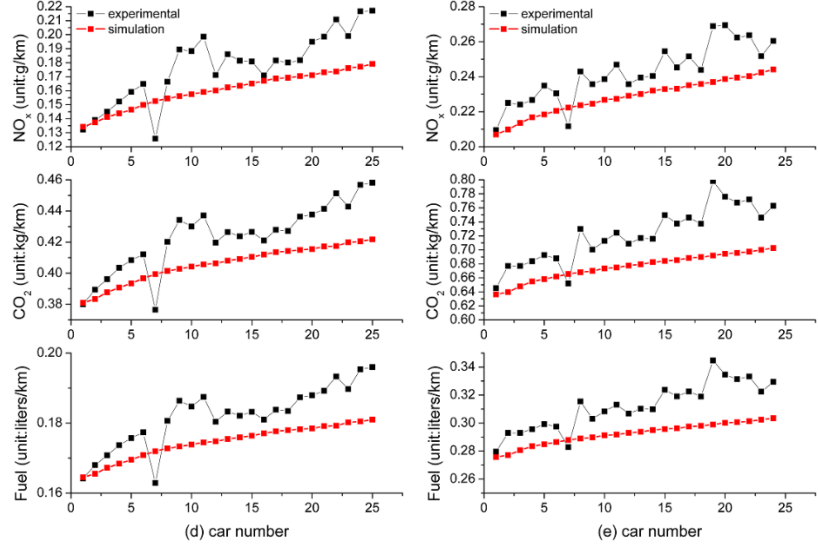

**Fig. 8.** The emission and fuel consumption of each car in the platoon. The symbol solid black lines are the experiment results and the red lines are the fitted lines. From (a) to (e), the speed of the leading car moves with $v_{\text{leading}}$ =50, 40, 30, 15, 7$km/h$ respectively. The car number 1 is the leading car.

## 4 The 2D-IIDM with memory effect and its simulation results

As Treiber and Helbing (2003) argued, driving behaviors might change according to the local surrounding. For example, after being stuck for some time in congested traffic, most drivers will increase their preferred netto bumper-to-bumper time gap to the preceding vehicle, which is named as memory effect.

Now we modify the 2D-IIDM by introducing the memory effect of drivers. Specifically, the adaptation of drivers to the surrounding traffic is assumed to be on time scales of a few minutes, which can be reflected by the following memory speed:

$$v_{n,\text{memo}} = \frac{1}{M} \sum_{i=1}^{M} v_n(t - i\Delta t) \tag{8}$$

where $\Delta t$=0.1$s$ is the time-step adopted by the car following models. The memory speed $v_{n,\text{memo}}$ is the average speed of vehicle $n$ in the past time interval [$t-M\Delta t$, $t-\Delta t$]. During the simulation, $M$=800 is used, which means that drivers will react according to their local surrounding in the past 80$s$.

Next, we show that the performance of 2D-IIDM can be improved by taking into account the memory effect. In 2D-IIDM, the two parameters $p_1$ and $p_2$ can be regarded as representing behavior changes since their values denote the changing frequencies of the desired time gap $T(t)$. Therefore, we assume the following relationships exist between memory speed $v_{n,\text{memo}}$ and the changing frequency parameters $p_1$ and $p_2$:

$$p_{1,n} = \max\left(\alpha_1 v_{n,\text{memo}} + \beta_1, \gamma_1\right) \tag{9}$$

$$p_{2,n} = \max\left(\alpha_2 v_{n,\text{memo}} + \beta_2, \gamma_2\right) \tag{10}$$

Through calibration, $\alpha_1=-0.00335 m^{-1}$, $\beta_1=0.0424 s^{-1}$, $\gamma_1=0.01 s^{-1}$, $\alpha_2=-0.00228 m^{-1}$, $\beta_2=0.0286 s^{-1}$, $\gamma_2=0.01 s^{-1}$. It means that the frequency of changing the desired gaps increases with the congestion experienced during the past 80$s$. i.e. $p_{1,n}$ and $p_{2,n}$ decrease with the memory speed $v_{n,\text{memo}}$.

The revised 2D-IIDM is named as 2D-IIDM with memory effect (abbreviated as 2D-IIDMM). Note that this reflects that drivers become somehow more agitated/aroused by changing their behavior more frequently. In this way, the model can be attributed to the framework of action-point models. In the original 2D-IIDM, the frequency of action points (where the behavior, i.e., the acceleration, changes abruptly) depends only mildly on the actual speed (or not at all if $p_{1,n}=p_{2,n}$). When memory is considered, this frequency depends additionally, and more strongly, on the past moving-average speed.

During the simulation, the parameters of 2D-IIDMM are set as the same as that of 2D-IIDM. The simulation results of 2D-IIDMM are presented in Fig.9-11, which significantly improve comparing with that of 2D-IIDM. In order to see the improvements more clearly, we have calculated the root-mean-square error (*RMSE*) and the Improvement Index (*IMI*) for 2D-IIDM and 2D-IIDMM as follows:

$$RMSE = \sqrt{\frac{1}{N}\sum_{n=1}^{N}\left(\frac{MOE_n^{\text{sim}} - MOE_n^{\text{experi}}}{MOE_n^{\text{experi}}}\right)^2} \tag{11}$$

$$IMI = \frac{RMSE_{\text{2D-IIDM}} - RMSE_{\text{2D-IIDMM}}}{RMSE_{\text{2D-IIDM}}} \tag{12}$$

where $N=25$ is the total number of cars in the car following platoon. Table 5 and 6 present the results of *RMSE* and *IMI* respectively, which demonstrate that with consideration of the memory effect, performance of 2D-IIDM does improve significantly. Table 4 shows that as in the experimental results, the correlations of emission and fuel consumption with both the speed oscillation and the standard deviation of acceleration are strong in the 2D-IIDMM.

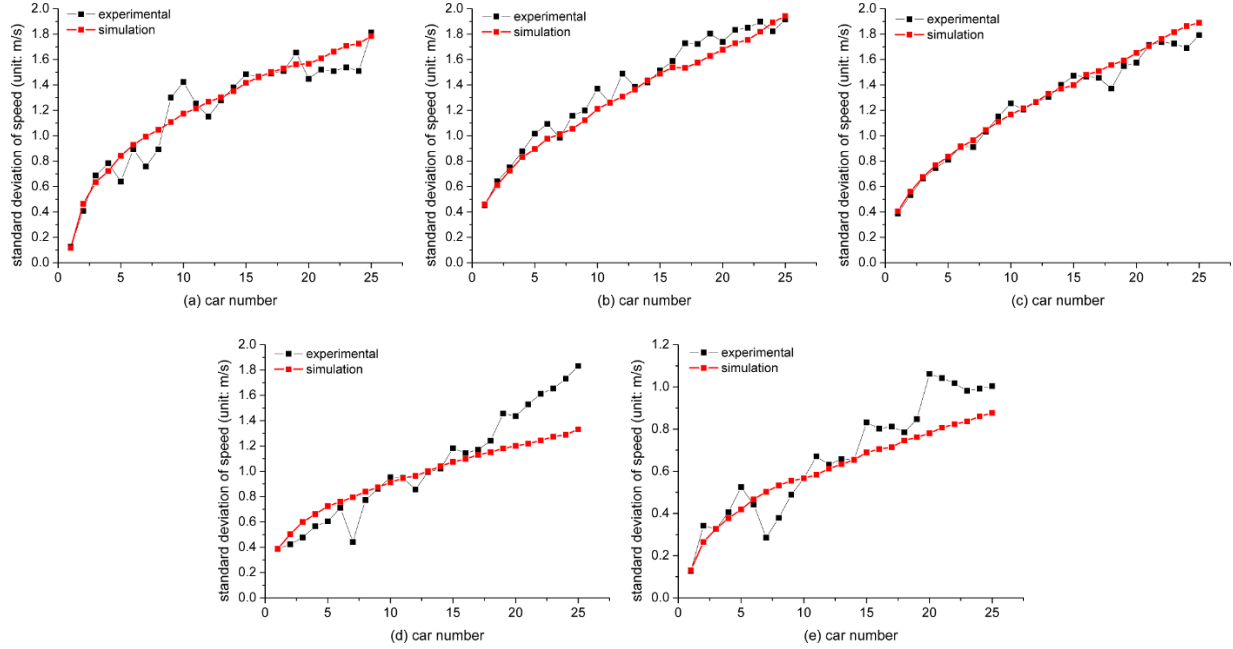

**Fig. 9.** The standard deviation of the time series of the speed of each car. The symbol solid black lines are the experiment results and the symbol solid red lines are the simulation results. From (a) to (e), the speed of the leading car moves with $v_{\text{leading}}$ =50, 40, 30, 15, 7$km/h$ respectively. The car number 1 is the leading car.

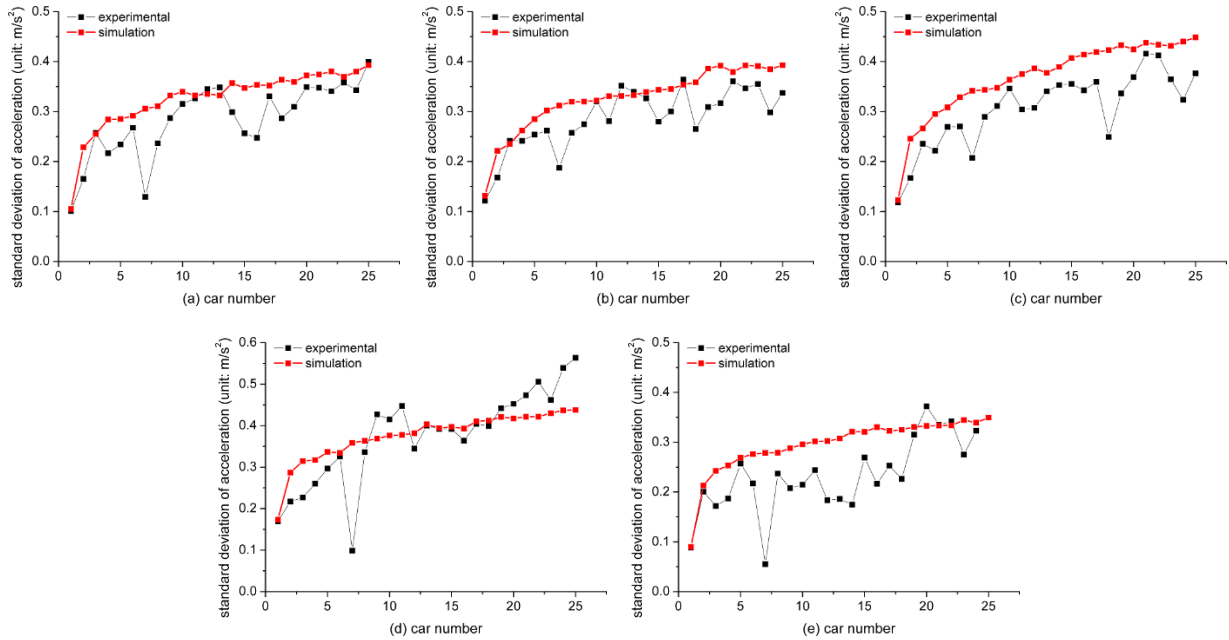

**Fig. 10.** The standard deviation of the time series of the acceleration of each car in the car following experiments. The symbol solid black lines are the experiment results and the red lines are the fitted lines. From (a) to (e), the speed of the leading car moves with $v_{\text{leading}}$ =50, 40, 30, 15, 7$km/h$ respectively. The car number 1 is the leading car.

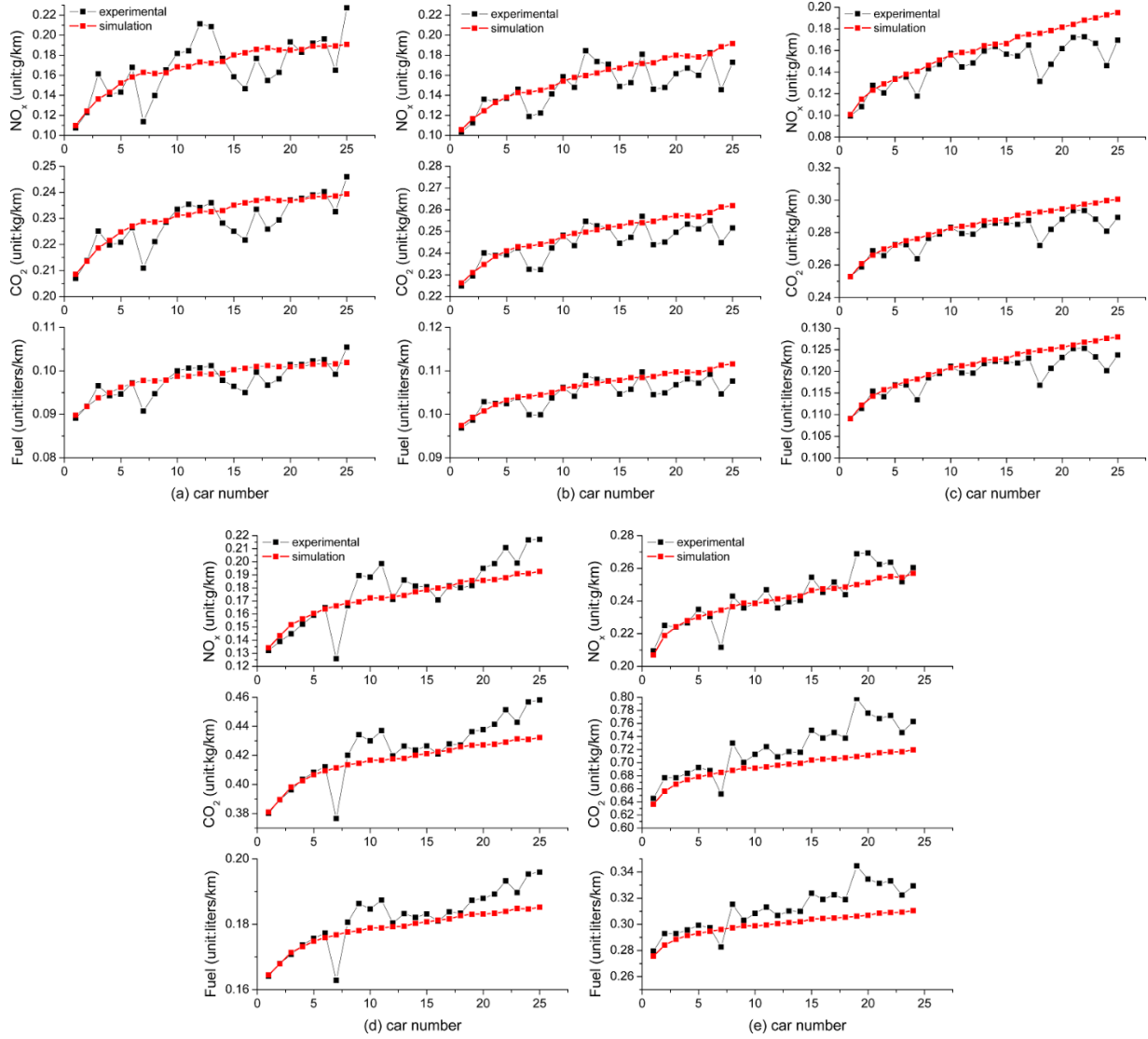

**Fig. 11.** The emission and fuel consumption of each car in the platoon. The symbol solid black lines are the experiment results and the red lines are the fitted lines. From (a) to (e), the speed of the leading car moves with $v_{\text{leading}}$ =50, 40, 30, 15, 7$km/h$ respectively. The car number 1 is the leading car.

**Table 5.** The root-mean-square error (*RMSE*)

| $v_{\text{leading}}$ | 50 | | 40 | | 30 | | 15 | | 7 | |
|---|---|---|---|---|---|---|---|---|---|---|
| (*km/h*) | 2D-IIDM | 2D-IIDMM | 2D-IIDM | 2D-IIDMM | 2D-IIDM | 2D-IIDMM | 2D-IIDM | 2D-IIDMM | 2D-IIDM | 2D-IIDMM |
| $\sigma_v$ (*km/h*) | 0.146 | 0.124 | 0.100 | 0.097 | 0.075 | 0.068 | 0.210 | 0.210 | 0.240 | 0.123 |
| $\sigma_a$ (*km/h*) | 0.249 | 0.059 | 0.238 | 0.053 | 0.238 | 0.073 | 0.434 | 0.074 | 0.237 | 0.105 |
| NO$_x$ (g/km) | 0.026 | 0.0222 | 0.0365 | 0.0174 | 0.0194 | 0.0186 | 0.0226 | 0.0148 | 0.0164 | 0.0084 |
| CO$_2$ (kg/km) | 0.0094 | 0.0066 | 0.0145 | 0.0069 | 0.0081 | 0.0078 | 0.0214 | 0.0139 | 0.0518 | 0.0367 |
| Fuel (liters/km) | 0.0037 | 0.0027 | 0.0058 | 0.0028 | 0.0031 | 0.0030 | 0.0089 | 0.006 | 0.0223 | 0.0160 |

**Table 6.** Improvement Index (*IMI*)

| $v_{leading}$ (km/h) | 50 | 40 | 30 | 15 | 7 |
|---|---|---|---|---|---|
| $\sigma_v$ (km/h) | 0.15 | 0.03 | 0.09 | 0.00 | 0.49 |
| $\sigma_a$ (km/h) | 0.76 | 0.78 | 0.69 | 0.83 | 0.56 |
| NO$_x$ (g/km) | 0.15 | 0.52 | 0.04 | 0.35 | 0.49 |
| CO$_2$ (kg/km) | 0.30 | 0.52 | 0.04 | 0.35 | 0.29 |
| Fuel (liters/km) | 0.27 | 0.52 | 0.03 | 0.33 | 0.28 |

## 5 Conclusion

Most of researches on traffic oscillations study the features such as period, propagation speed, or whether the oscillation grows or decays. The growth pattern of oscillation has seldom been investigated. This is perhaps due to the scarcity of trajectory data. Recently, Jiang et al. (2014, 2015) have conducted an experimental study of car following behaviors in a 25-car-platoon on an open road section. They found that the speed oscillation of each car increases in a concave way along the platoon.

This paper makes a further analysis of the traffic oscillation features. We have studied the standard deviation of acceleration, fuel consumption and emission in the car-following platoon. It has been found that: (1) the three indices increase along the platoon in a concave way, which is a common feature as the growth pattern of the speed oscillation; (2) As average speed of the platoon $\bar{v}$ declines, the growth pattern of emission and fuel consumption is more and more close to the linear way; (3) Emission of CO$_2$ and NO$_x$ exhibit different dependence on $\bar{v}$. Roughly speaking, the emission and fuel consumption of each vehicle decrease remarkably when $\bar{v}$ increases from low value; However, when $\bar{v}$ reaches 30 *km/h*, the change of emission and fuel consumption with $\bar{v}$ is not so significant; (4) the correlations of emission and fuel consumption with both the standard deviation of acceleration and the speed oscillation are strong. Simulations show that with the memory effect of drivers taken into account, the 2D-IIDMM is able to reproduce the common feature of traffic oscillation evolution quite well.

In our future work, we plan to conduct the following researches: (1) utilize the experimental results to examine other car following models; (2) analyze the standard deviation of acceleration, emission and fuel consumption in the empirical data; (3) carry out larger-scale car following experiments on longer road section with larger platoon size and higher speed.

**Acknowledgements:**

This work was supported by the National Basic Research Program of China under Grant No. 2012CB725400. JFT was supported by the National Natural Science Foundation of China (Grant No. 71401120). BJ was supported by the National Natural Science Foundation of China (Grant No. 71222101). RJ was supported by the Natural Science Foundation of China (Grant Nos. 11422221 and 71371175). SFM was supported by the National Natural Science Foundation of China (Grant No. 71271150 and 71431005). WYZ was supported by China Postdoctoral Science Foundation (Grant No. 2015M580973). Correspondence and requests for materials should be addressed to BJ and RJ.